\documentclass[9pt,twocolumn,twoside]{osajnl}

\journal{ao} 

\usepackage{subfigure}
\setboolean{shortarticle}{true}

\title{High fidelity single-pixel imaging}

\author[1,2]{Chao Deng}
\author[3]{Xuemei Hu}
\author[1]{Xiaoxu Li}
\author[1]{Jinli Suo}
\author[2]{Zhili Zhang}
\author[1]{Qionghai Dai}

\affil[1]{Department of Automation, Tsinghua University, Beijing, 100084, China}
\affil[2]{High-Tech Institute of Xi’an, Xi’an, 710025, China}
\affil[3]{School of Electronic Science and Engineering, Nanjing University, Nanjing, 210000, China}

\begin{abstract}
Single-pixel imaging (SPI) is an emerging technique which has attracts wide attention in various research fields. However, restricted by the low reconstruction quality and large amount of measurements, the practical application is still in its infancy.  Inspired by the fact that natural scenes exhibit unique degenerate structures in the low dimensional subspace, we  propose to take advantage of the local prior in convolutional sparse coding to implement high fidelity single-pixel imaging. Specifically, by statistically learning strategy, the target scene can be sparse represented on an over-complete dictionary.  The dictionary is composed of various basis learned from a natural image database. We introduce the above local prior into conventional SPI  framework to promote the final reconstruction quality.  Experiments both on synthetic data and real captured data demonstrate  that our method can achieve better reconstruction from the same measurements, and  thus consequently reduce the number of required measurements for same reconstruction quality.
\end{abstract}

\setboolean{displaycopyright}{true}

\begin{document}

\maketitle

\section{Introduction}
Single-pixel imaging (SPI) is a novel imaging scheme which correlates two beams non-locally. These two light beams travel differently: one beam interacts with object and is recorded by a bucket detector, the other is directly collected by a spatially resolved detector.  The object can be reconstructed by correlating the output of these two beams. At the beginning, SPI was considered as the unique mechanism of quantum entangle photons \cite{pittman1995optical}, which was referred as ghost imaging (GI). Soon after,  classic thermal light was also proved to be successful  under the same configuration \cite{bennink2002two,gatti2004ghost,valencia2005two}. Programmable illumination makes SPI more flexible and ready to put into practical applications, such as fluorescence imaging \cite{tian2011fluorescence}, remote sensing \cite{zhao2012ghost}, 3D reconstruction \cite{sun20133d}, optical encryption \cite{clemente2010optical,chen2013ghost}, and object tracking \cite{magana2013compressive,li2014ghost}.

So far, three typical categories of reconstruction algorithms for SPI has been proposed: linear correlation method \cite{bromberg2009ghost,gong2010method,ferri2010differential,sun2012normalized}, the alternating projection (AP) method \cite{guo2016multilayer,wang2015gerchberg,deng2018single} and compressive sensing (CS) based method \cite{katz2009compressive,abetamann2013compressive}. The correlation-based methods  restore the specific object through second order or higher order correlations, which suffer from low reconstruction  quality and  large amount of measurements. The AP algorithm incorporates the constraints from the patterned illumination and  correlated measurements  in spatial and Fourier domain alternatively. It typically iterates $100\sim200$ rounds (each round includes all the iteration over each measurement) until the final convergence. In contrast, the compressed sensing-based methods are more efficient which need far lower measurements with faster convergence compared with the former two.

The compressed sensing algorithm in SPI introduces the image priors into the under-determined linear system to  decrease the solution space. Two widely used priors are sparse representation prior and the total variation (TV) regularization  prior \cite{bian2018experimental}. The former states that natural images can be sparse represented in some orthogonal basis as DCT and wavelet \cite{yu2014adaptive,gong2015high}, and the latter considers the gradient integral as statistically low. Such global priors constraint all the natural images  from the general perspective and improve the final reconstruction.  Except for global prior, the natural images of specific class exhibit unique degenerate structure, which can be sparse denoted in the low-dimensional self-characteristic subspace. For example, image statistics suggest that the image patches can be well represented on an over-complete dictionary \cite{yang2010image,wright2010sparse,bao2016dictionary,hu2015patch}, which is termed as the sparse coding (SC). Local prior as such in sparse coding has been widely used and achieved  state-of-the-art performances  in computer vision, such as  denoising, deblurring, impainting, super resolution imaging and machine learning.

Sparse coding aims to construct an over-complete dictionary in which a sparse linear combination of atoms can well approximate the original image. The atoms of over-complete dictionary  are  intrinsic structures  learned from thousands of images. Patch-based SC considers overlapped patches as sub-elements and processes them separately. Each patch can be sparse represented and the whole image is reconstructed by the average of these patches. Although the patch-based method can reduce the calculation size in optimization and achieve high-quality representation, two limitations arising in this method suppress the final reconstruction in various task.  First, since  the simple variance of intrinsic structure (such as shift or rotation) is indistinguishable due to the independent mechanism in patch-based SC,  the learned dictionary  is  highly redundant. Second, the overlap-averaging mechanism leads to the inconsistency of overlapped patches \cite{gu2015convolutional}.  As opposed to the patched-based SC, the convolutional sparse coding (CSC)  decomposes the whole image into some sparse feature maps  thus avoid the prior consistency  of learned atoms and inconsistency of overlapped patches. The CSC can model the image more explicitly than patched-based SC through convolution operator.

In this paper, we propose to incorporate the local prior in convolutional sparse coding with the global prior to implement high fidelity single-pixel imaging.
Our framework can  greatly improve the reconstruction in non-spatially resolved imaging technique compared with current state-of-the-art methods, and have further promotion  in practical application.
The remaining part of the paper is organized as follows: In section \ref{framework}, we  introduce our modeling and derivative process mathematically. In Section \ref{simulation},  we evaluate the performance of our method  both on simulation and experiment. Finally in section 4, we summarise the disadvantages and limitations of  our  work.

\section{Method}
\label{framework}

To fully exploit the image prior both globally and locally, we incorporate these two priors together  for high fidelity reconstruction.
Mathematically, we can learn the kernels of CSC from large numbers of nature images by the following optimization \cite{bristow2013fast,heide2015fast}
\begin{eqnarray}\label{CSC}
\mathop{\arg\min}_{{\bf d, s}}&&\sum_{j=1}^J\frac{1}{2} \|{\bf x}^j-\sum_{k=1}^K{\bf d}_k\ast {\bf s}_k^j\|_2^2+\beta\sum_{k=1}^K\|{\bf s}_k^j\|_1\nonumber\\
\text{s.t.}&&\|{\bf d}_k\|_2^2\leq1 \quad \forall  k\in\{1, \cdots, K\},
\end{eqnarray}
where  each image ${\bf x}^j$  can be represented as the summation of  vectorized  2D kernels ${\bf d}_k$ convolved with corresponding sparse feature maps ${\bf s}_k^j$. Here ${\bf x}^j$ and ${\bf s}_k^j$ are both  vectorized  form. $\beta$ weights the $\ell_1$ penalty, and we set $\beta=1$ for its good tradeoff between sparsity and data fitting as in \cite{heide2015fast}.
The operation $\ast$ denotes the 2D convolution defined on the vectorized inputs.  We use the fast and flexible algorithm proposed in \cite{heide2015fast}  to obtain the over-complete dictionary of convolve sparse coding.
\begin{figure}[htb]
  \centering
  \includegraphics[width=1.0\linewidth]{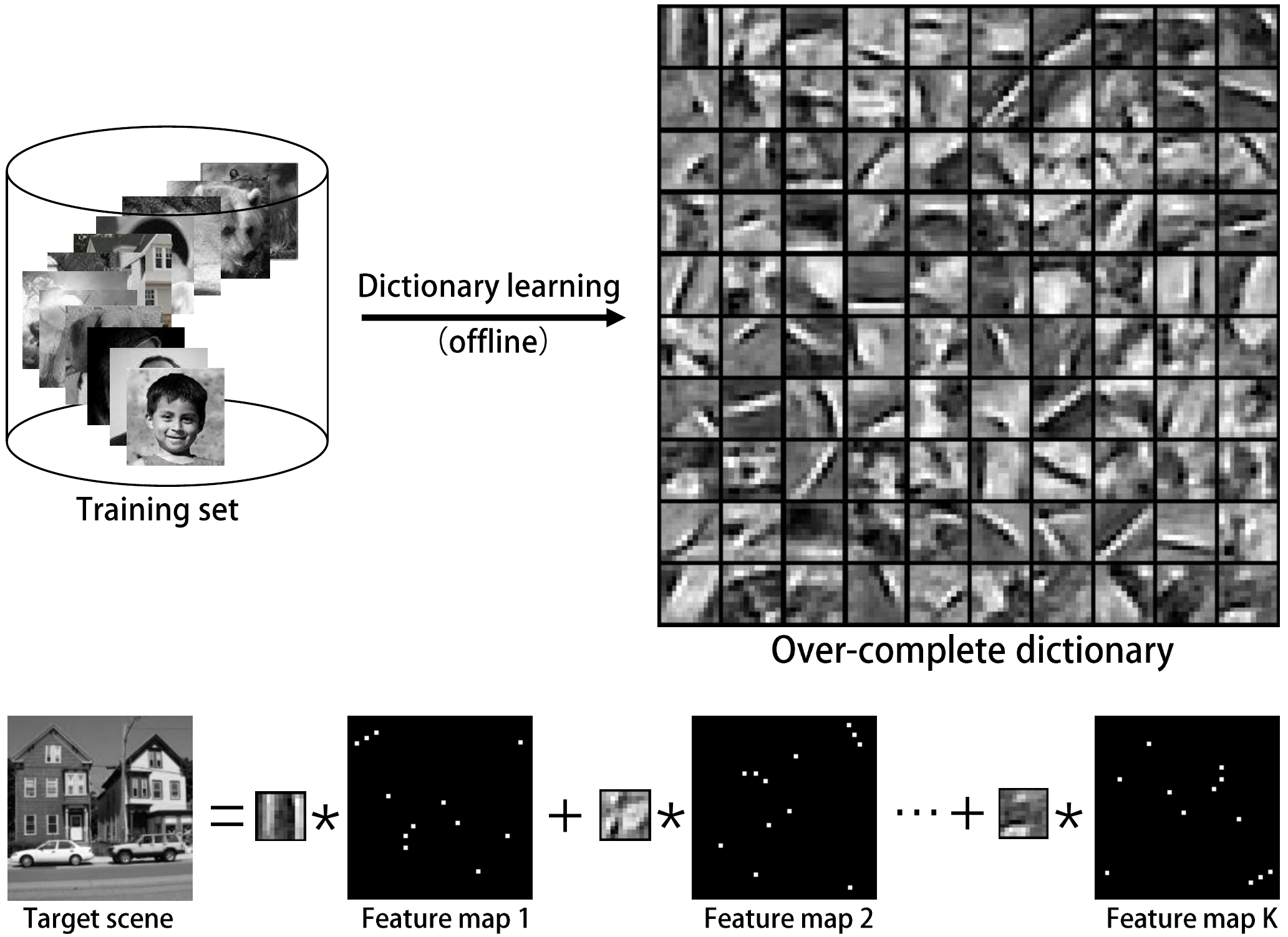}\\
  \caption{The overview of our model. The upper part describes the learning process of convolutional sparse coding.
  With the learned dictionary, the target scene can be decomposed as summation of kernels  convolved with corresponding sparse feature maps.
     }\label{sparseCoding}
\end{figure}

After the kernel learning as shown in Fig.~\ref{sparseCoding}, the over-complete dictionary can be utilized to decomposed the target image. Introducing this local prior into SPI  reconstruction, the
optimization can be modeled as
\begin{eqnarray}\label{CSCTV}
  \mathop{\arg\min}_{{\bf s}_k}&& \|{\bf \Psi x}\|_1+\lambda\sum_{k=1}^K\|{\bf s}_k\|_1\nonumber\\
  \text{s.t.}&& {\bf y=\Phi x} \nonumber\\
             && {\bf x}=\sum_{k=1}^K {\bf s_k \ast d_k}.
\end{eqnarray}
Here ${\bf \Psi}$ is the transformation matrix of desired domain as DCT or TV, ${\bf \Phi}$ is the sampling matrix with each row denoting the vectorized illumination pattern, ${\bf y}$ is the single pixel measurement and $\lambda$ is the penalty parameter. The first term in the objective function is  the global constraint defined in the DCT or TV domain, and the second term local constraint on the learned dictionary.

For simplicity, we remove the intermediate variable ${\bf x}$ and make the simple substitutions in  Eq.~\ref{CSCTV}, it can be rewritten as
\begin{eqnarray}\label{CSCTV2}                                                                
  \mathop{\arg\min}_{{\bf s}_k}&& \|{\bf v}\|_1+\lambda\sum_{k=1}^K\|{\bf s}_k\|_1\nonumber\\
  \text{s.t.}&& {\bf y=\Phi}\cdot \sum_{k=1}^K {\bf u}_k\nonumber\\
             && {\bf v}={\bf \Psi}\cdot \sum_{k=1}^K{\bf u}_k\nonumber\\
             && {\bf u}_k={\bf s}_k \ast {\bf d}_k.
\end{eqnarray}
 We solve above optimization by alternating direction method of multipliers (ADMM) algorithm \cite{boyd2011distributed}, and the target scene ${\bf x}$ can  be obtained by optimum ${\bf s}_k^\ast$.                    As for convolution operator  in Eq.~(\ref{CSCTV2}), we first  transform ${\bf s}_k$ and ${\bf d}_k$ into respective Fourier space, and then implement Hadamard product.
 After  Hadamard product in Fourier space, we inverse  transform back to spatial domain. These calculation is  faster than convolution operator since Fourier transform and Hadamard  product
 are both simple matrix multiplication. 


\section{Experiments on synthetic data}
\label{simulation}

To evaluate the performance of our method, we  implement a series of  simulations on the synthetic data. We first learn the over-complete dictionary of CSC by 20
randomly chosen images from the dataset built by the Stanford Vision Lab \cite{russakovsky2015imagenet}. To balance the computational load and over-completeness, each kernel of the learned dictionary is set to $11\times11$, and we have 100 kernels in total (The learned dictionary is shown in Fig.~\ref{sparseCoding}). We adopt the same dictionary size as in \cite{heide2015fast} to achieve good performance for general natural scenes.
Then we conduct the CSC-based SPI framework using the ADMM optimization .  The imaging size of target scene and modulation patterns are both $128\times128$. All the test images are chosen out of training set. The measurement is generated  by their inner product.  We integrate CSC  together with  two aforementioned global priors as in TV and DCT domain, since these two constraints can achieve state-of-the-art performance in SPI.

\begin{figure}[htb]
  \centering
  \includegraphics[width=1.0\linewidth]{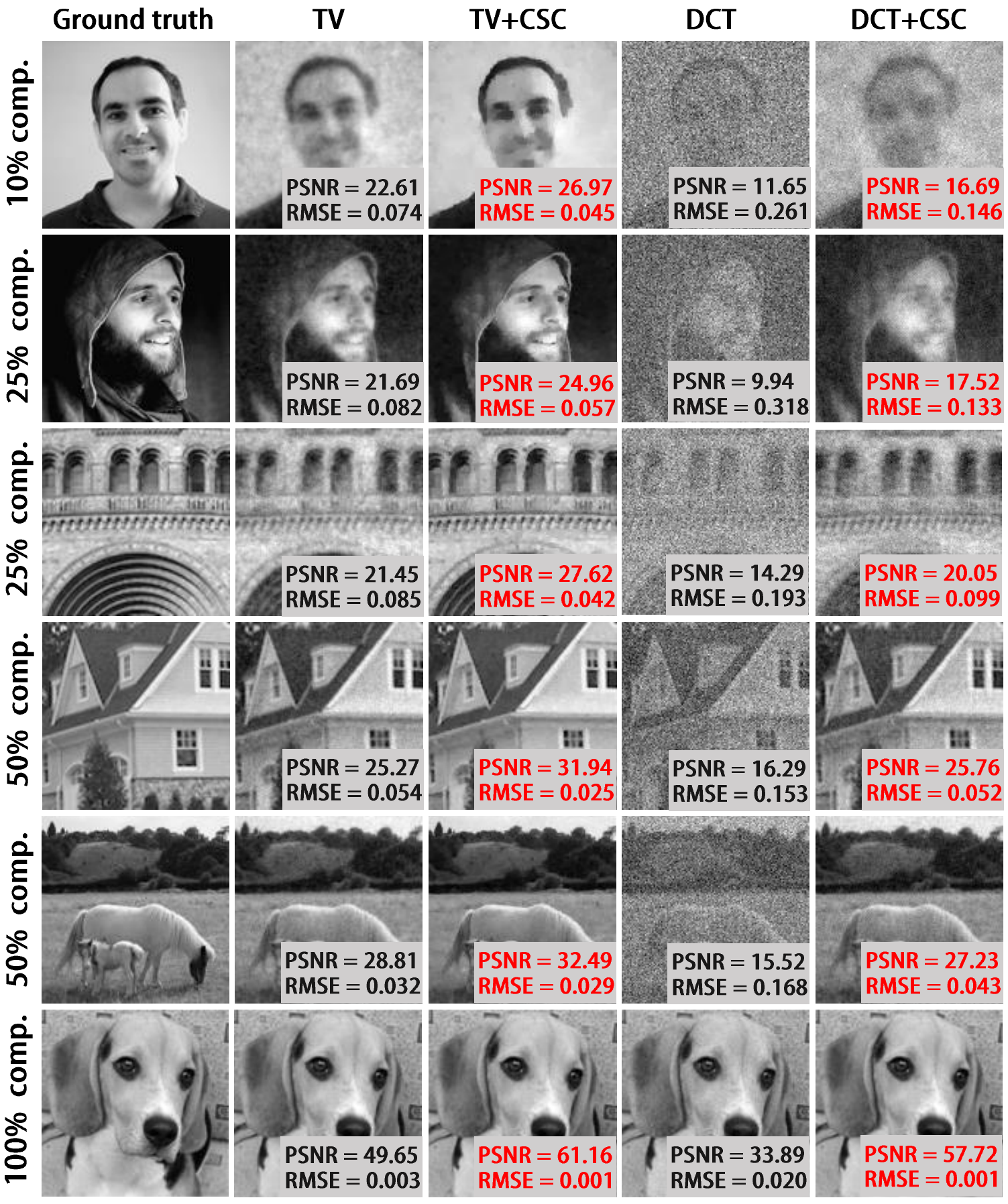}\\
  \caption{The reconstruction comparison of four different combinations: TV, TV+CSC, DCT and DCT+CSC, under different compression ratio in terms of PSNR and RMSE. }\label{CSCpromotion}
\end{figure}

To show the performance of CSC prior, we first conduct a simulation on several images under different compression ratio (i.e., pattern number divided by pixel number).   We compare the reconstruction of two different groups: TV and TV+CSC, DCT and DCT+CSC, under the compression of 10\%, 25\%, 50\% and 100\%. The simulation result is shown in Fig.~\ref{CSCpromotion}. The reconstruction using TV prior exhibits higher quality than DCT prior since all the test images are with non-periodic structures. In terms of peak-to-signal ratio (PSNR) and root mean square error (RMSE), the introduction of CSC prior can significantly improve the SPI reconstruction. Specifically, in this simulation,  the promotion of PSNR is about 3.27 dB $\sim$ 11.51 dB due to the CSC introduction into TV, and it turns 5.04 dB $\sim$ 23.33 dB because of the CSC prior into DCT. Except for the PSNR, the improvement in RMSE is also significant. Different compression ratios and different images may vary in the promotion.  The combination  both globally and locally  further reduces the solution space and improves the reconstruction accuracy.

\begin{figure}[htb]
  \centering
  \subfigure[]{
  \includegraphics[width=1\linewidth]{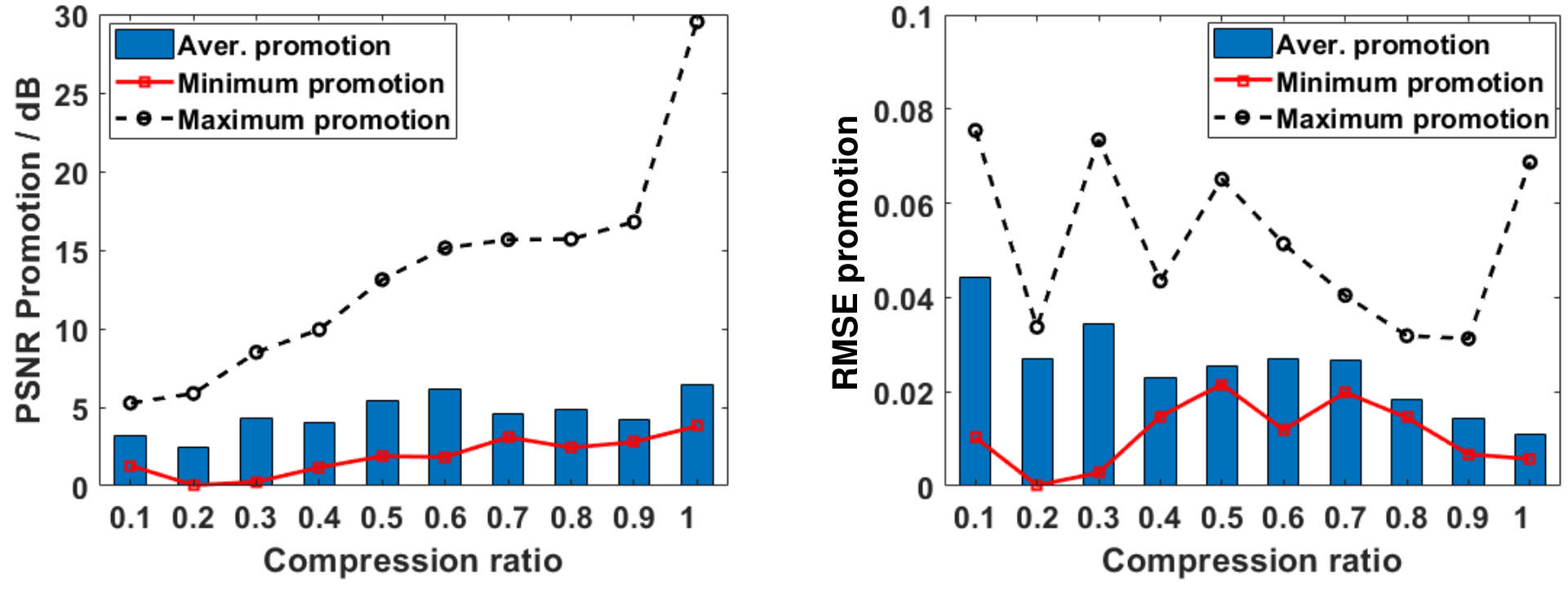}}\\
  \subfigure[]{
  \includegraphics[width=1\linewidth]{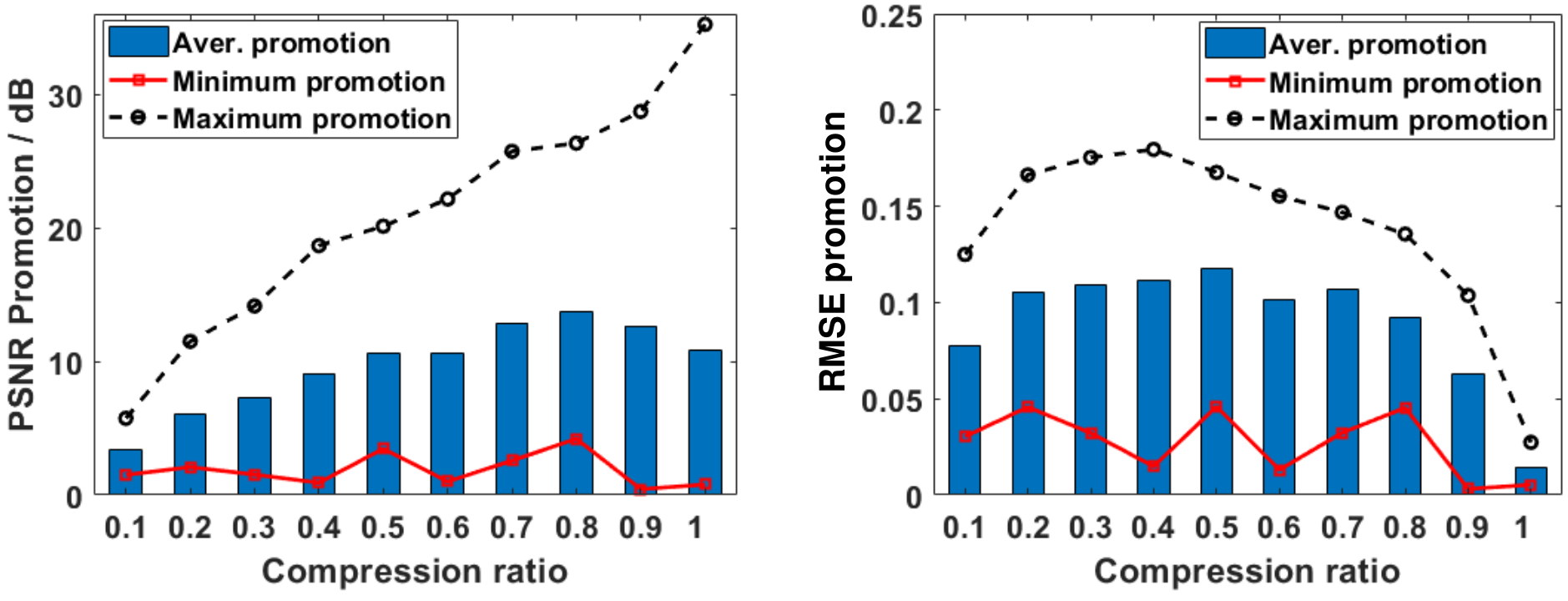}}\\
  \caption{Promotion of CSC introduced into TV (a) and DCT (b) in terms of PSNR and RMSE under different compression ratios.}\label{averagePromotion}
\end{figure}

For a  better  evaluation of the CSC prior, we investigate the average promotion on the image set under different compression ratio, which is ranging from 0.1 to 1.0 with the interval of 0.1. We choose 12 images  from the database as the test set, which  is composed of four different groups: person, animal, building and scenery, with 3 images in each group. We plot the average value and the extremum in Fig.~\ref{averagePromotion}.  The simulation result suggests that the averaged improvement of CSC prior introduced  into TV prior is about 2.5 $\sim$ 6.4 dB (PSNR) and 0.011$\sim$0.044 (RMSE), while the promotion of CSC into DCT is 3.4 $\sim$13.7 dB (PSNR)  and 0.015$\sim$0.118 (RMSE). The advantage of introducing CSC is more distinct in DCT than TV. Larger compression ratio may be more likely to exhibit more marked PSNR promotion both in TV and DCT. The improvement of CSC introduction exhibit severe  polarization on different images.  The small value of minimum promotion curve close to 0 in PSNR and RMSE may caused by the untrained structure in training set. We can include more images into training set and increase the number of kernels  to learn a more sufficient and effective over-complete dictionary. In sum, the simulation result in Fig.~\ref{averagePromotion} demonstrates that high fidelity image quality in SPI can achieved when we add the local prior of intrinsic structure into reconstruction.

\begin{figure}[b]
  \centering
  \subfigure[]{
  \includegraphics[width=0.48\linewidth]{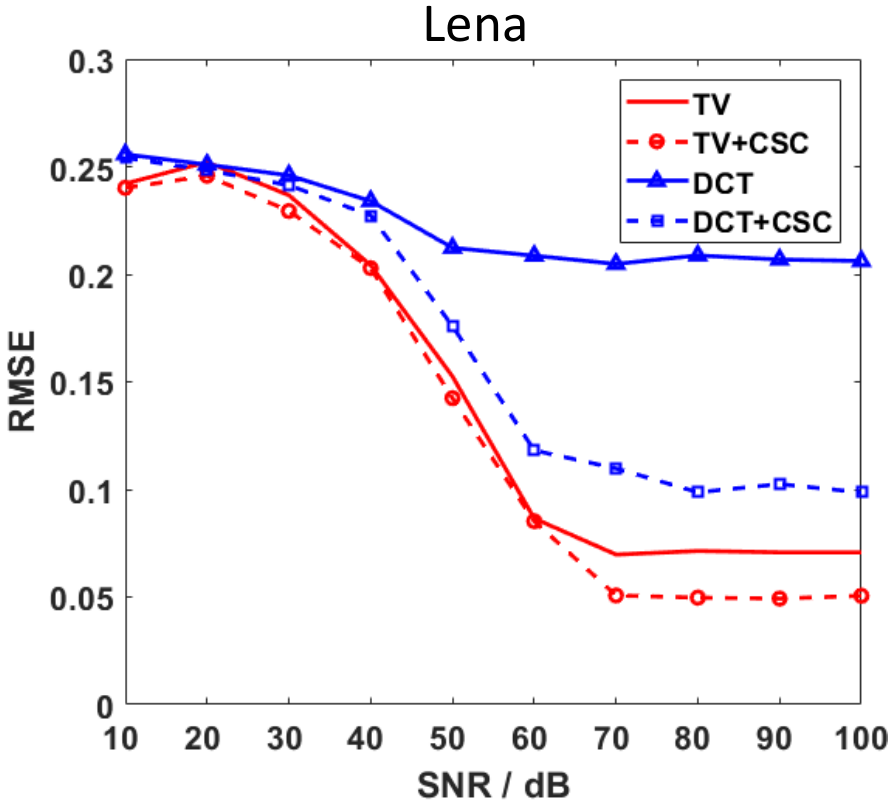}}
  \subfigure[]{
  \includegraphics[width=0.48\linewidth]{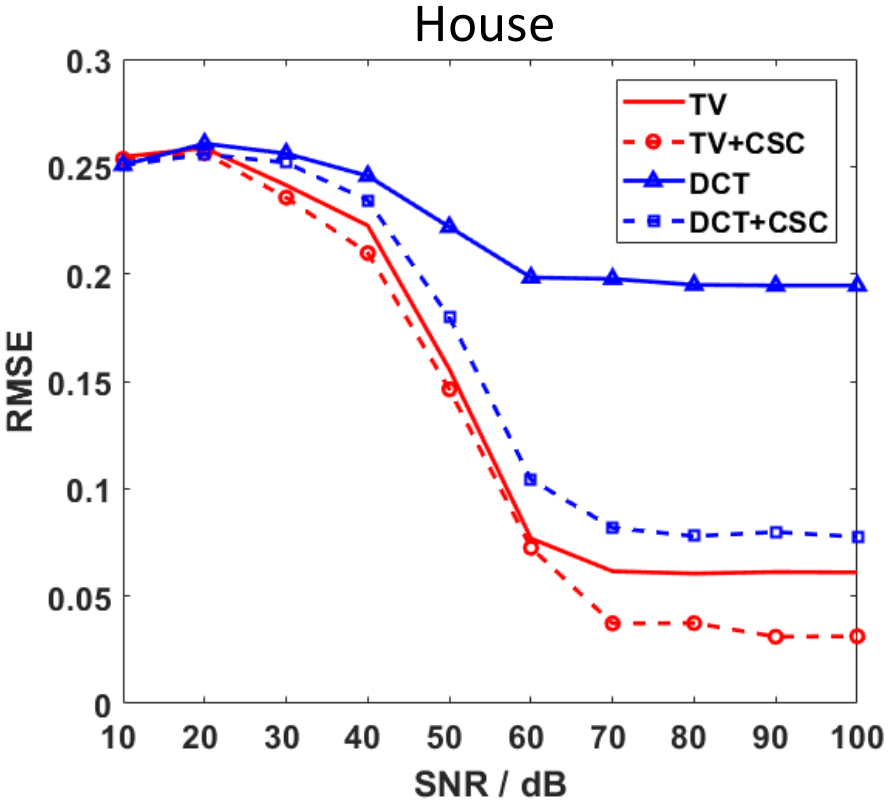}}\\
  \caption{The robust to imaging noise  among four different algorithms of (a): ``Lena'' and (b): ``House''.}\label{noise}
\end{figure}

The noise in the imaging process such as the sensor readout noise or intensity fluctuation of light source  is inevitable in actual SPI experiment, so here we test the  anti-noise capability of four aforementioned  algorithms. We model the imaging noise as superimposing Gaussian white noise with the signal-to-noise ratio (SNR) ranging from 10 dB to 100 dB, and choose the RMSE as the evaluation metric.  Without loss of  generality, we specify the compression ratio as 0.25 in simulation. By comparing the RMSE scores of reconstruction under  four different algorithms, the simulation  results of two classic images ``Lena'' and ``House'' are shown in Fig.~\ref{noise}. Overall, the reconstruction of all the algorithms gets better as the noise level decreases, and becomes stable when the  SNR exceeds 70 dB.  As aforementioned, since the DCT algorithms prefer periodic structures, TV related methods have better performance than DCT related ones, especially under higher SNR. The gap before and after CSC introduction is increasing as the noise level decreases both in TV and DCT. This result guides us in actual experiment a lot: by minimising  the imaging noise, we could achieve higher promotion  through introducing additional CSC prior.

\section{Experiments on real captured data}

\begin{figure}[htb]
  \centering
  \includegraphics[width=1.0\linewidth]{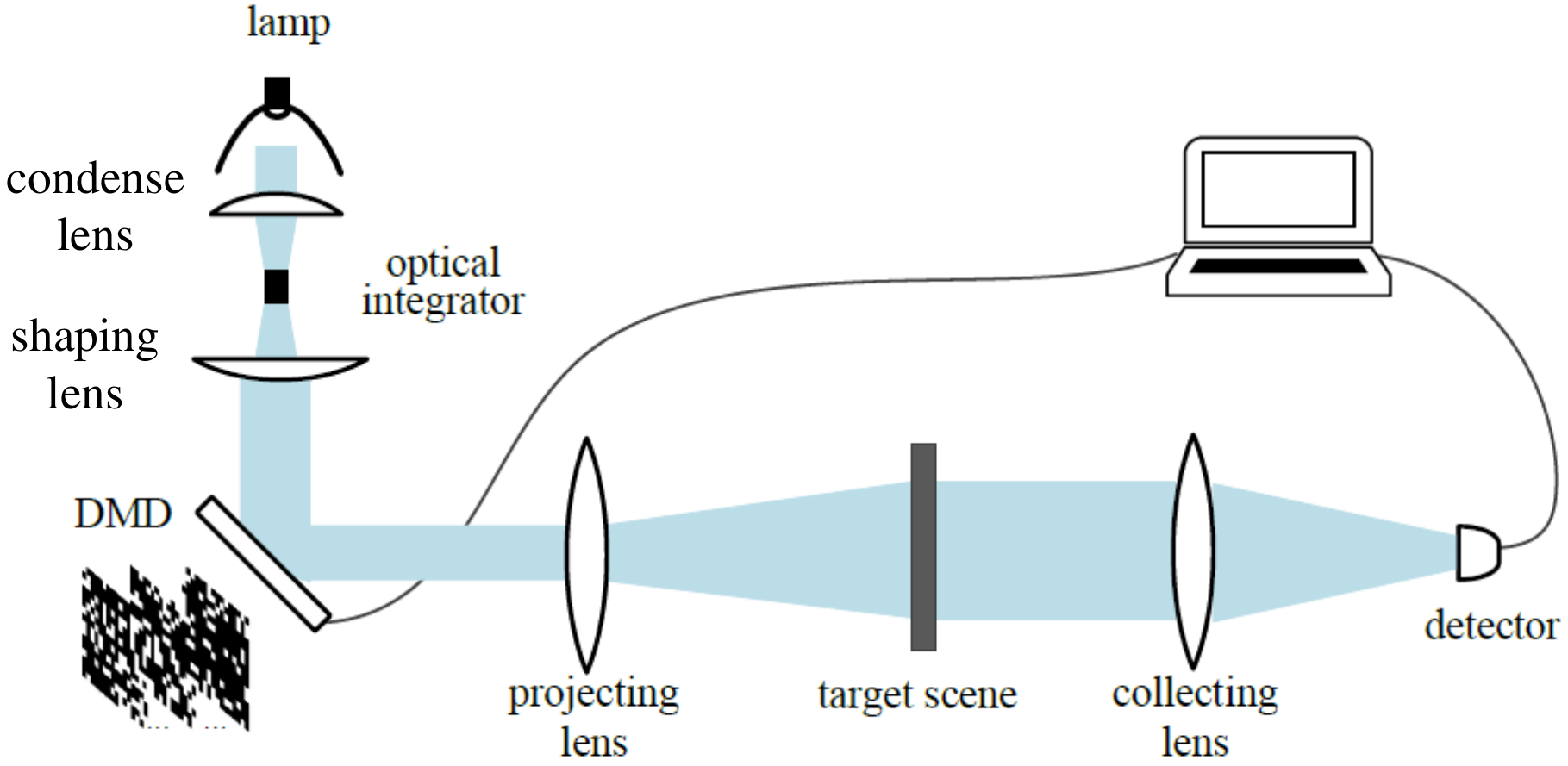}\\
  \caption{The schematic diagram of our experimental setup.  The high pressure Mercury lamp (Philips, 200w) is used as the light source. DMD: digital micromirror divece (Texas Instrument DLP \circledR DiscoveryTM4100, .7XGA). Detector: Thorlabs DET100 Silicon photodiode (integration time: 0.625ns). }\label{lightpath}
\end{figure}

Finally, we experimentally demonstrate the performance of our method by the captured data of a prototype system. The scheme of our experimental setup is shown in Fig.~\ref{lightpath}. The light emitted from the halogen lamp is first collimated by a collimator composed of a condense lens, an optical integrator and a shaping lens. After collimation, a digital micromirror device (DMD) modulates  the incoming light with random patterns. The outgoing light from the DMD is expanded by the projecting lens for scene illumination. After interaction with target scene, the outgoing photons  are converged by a  collecting lens and finally captured by a bucket detector.

\begin{figure}[htb]
  \centering
  \includegraphics[width=1\linewidth]{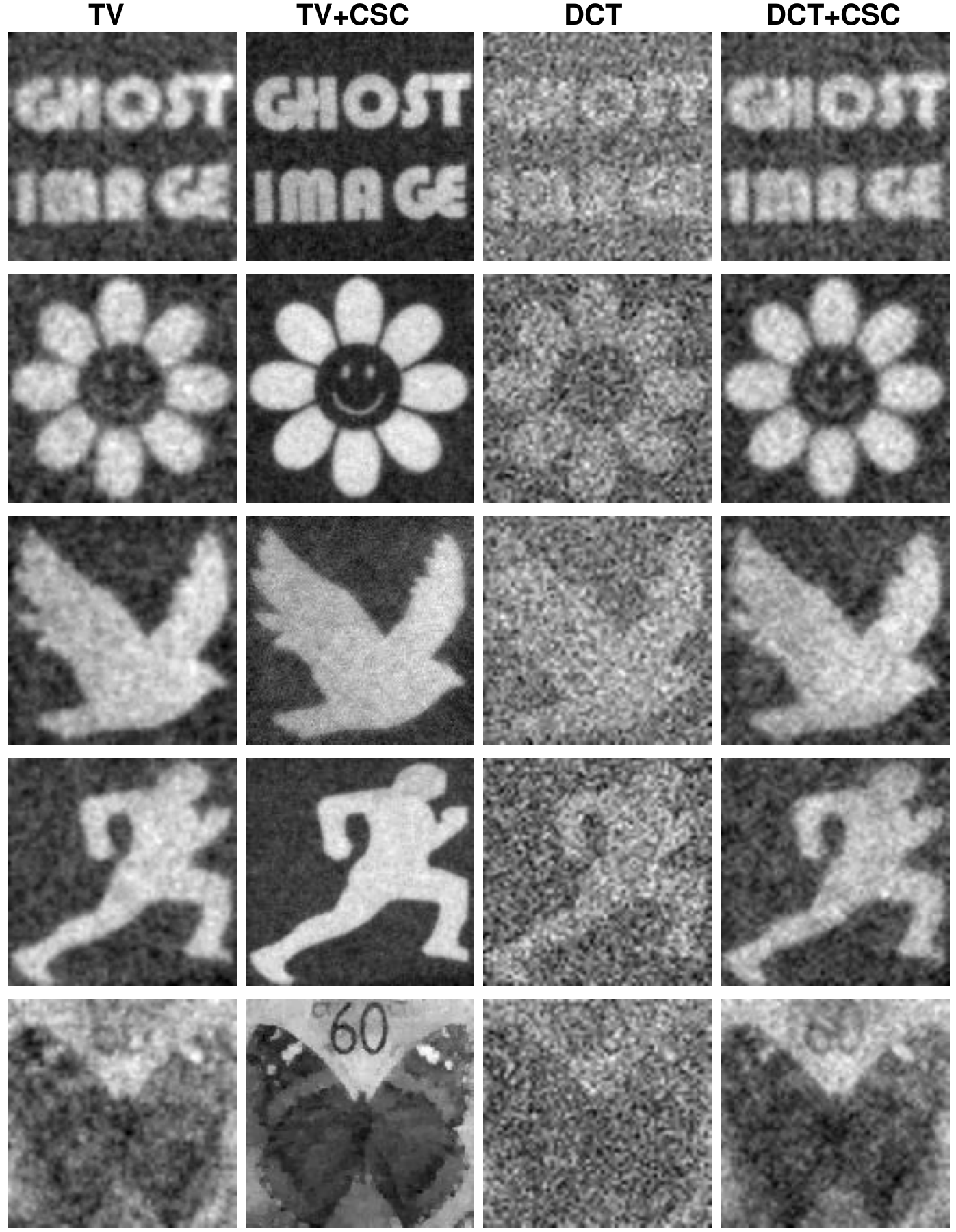}\\
  \caption{The reconstruction of the experiment on five different scenes. The compression ratio is 0.25, and each measurement is averaged under 200 samplings.}\label{experiment}
\end{figure}

In implementation, the pixel resolution  is $128\times128$, consistent with the simulation. We set the compression ratio as 0.25 and capture approximately 4100 measurements. During imaging process, several factors might influence the final reconstruction. For example, the fluctuation of light source and background light, the instability of detector sensitivity, and minor vibration of the light path. To effectively  suppress the imaging noise, we conduct each measurement by averaging over 200 samplings, and maximize the amplification gain of the bucket detector to minimize the detected noise. As can be seen from the Fig.~\ref{experiment}, although there still exist a bit artifacts in the reconstruction by introducing the local prior of CSC,   the reconstruction of our method  perform significant superiority to conventional ones. By additional local prior constraint, the reconstruction of 0.25 compression ratio under experimental condition is decent. The noisy background is suppressed and local details is clearer. For example, the smiling mouth in the image ``flower'' exhibits  clearer edge after CSC introduction; the number ``60'' of the image ``Butterfly'' can be obviously resolved in TV+CSC. In sum, through imposing sparse constraints both globally and locally on the SPI framework,  the reconstruction quality  could be improved significantly.

\section{Conclusion}

In conclusion, this paper propose a high fidelity single-pixel imaging scheme by introducing the local prior of convolutional sparse coding into  global prior as TV and DCT.   By imposing the combinational constraints on the SPI framework, the reconstruction exhibit significant promotion and thus can decrease the numbers of measurements in practical application.  The simulations on the synthetic data tell that the averaged promotion of different compression ratio is approximately $2.5\sim6.4$ dB in TV and $3.4\sim 13.7$ dB in DCT. Moreover, the experiment  also demonstrate the effectiveness of our method.

In terms of computational load, the optimization process of our method  would not increase the calculation complexity, compared with conventional compressive sensing based method. The over-complete dictionary can be learned before hand,  and the convolutional operator  in the optimization could be transformed into frequency domain for matrix multiplication.

Our method can be accomplished  without  hardware modification based on the conventional SPI scheme. In implementation, considering only 20 natural images is used for training the dictionary,  we  could obtain higher representative precision through training more natural images or training specific class of natural scenes for the certain target scene. To sum up, our method has the potential to broaden the practical application of single-pixel imaging.

\section*{Funding Information}

National key foundation for exploring scientific instrument of China (2013YQ140517);
National Science Foundation of China (61327902, 61631009).

%
%
%
%




\end{document}